\documentclass[aps,pre,twocolumn,superscriptaddress]{revtex4}
\usepackage{graphics,color}
\usepackage[dvips]{graphicx}
\usepackage{times}
\usepackage{multirow}
\usepackage{float}
\usepackage{blkarray}
\usepackage{lipsum}
\begin{document}

\title{Evolutionary games on simplicial complexes}

    \author{H. Guo}
 	\affiliation{School of Mechanical Engineering, Northwestern Polytechnical University, Xi'an 710072, China}
    \affiliation{School of Artificial Intelligence, Optics and Electronics (iOPEN), Northwestern Polytechnical University, Xi'an 710072, China}
	
    \author{D. Jia}
    \affiliation{School of Mechanical Engineering, Northwestern Polytechnical University, Xi'an 710072, China}
    \affiliation{School of Artificial Intelligence, Optics and Electronics (iOPEN), Northwestern Polytechnical University, Xi'an 710072, China}

    \author{I. Sendi\~na-Nadal}
	\affiliation{Universidad Rey Juan Carlos, Calle Tulip\'an s/n, 28933   M\'ostoles, Madrid, Spain}
    \affiliation{Center for Biomedical Technology, Universidad Polit\'ecnica de Madrid, 28223 Pozuelo de Alarc\'on, Madrid, Spain}

    \author{M. Zhang}
	\affiliation{School of Ecology and Environment, Northwestern Polytechnical University, Xi'an 710072, China}
    	
	\author{Z. Wang}
	\affiliation{School of Mechanical Engineering, Northwestern Polytechnical University, Xi'an 710072, China}
	\affiliation{Center for OPTical IMagery Analysis and Learning (OPTIMAL), Northwestern Polytechnical University, Xi'an 710072, China}
	
	\author{X. Li$^{+,}$}
	\affiliation{Center for OPTical IMagery Analysis and Learning (OPTIMAL), Northwestern Polytechnical University, Xi'an 710072, China}
	
    \author{K. Alfaro-Bittner}
	\affiliation{Unmanned Systems Research Institute, Northwestern Polytechnical University, Xi'an 710072, China}
	\affiliation{Departamento de F\'isica, Universidad T\'ecnica Federico Santa Mar\'ia, Av. Espa\~na 1680, Casilla 110V, Valpara\'iso, Chile}

	\author{Y. Moreno}
	\affiliation{ISI Foundation, Turin, Italy}
    \affiliation{Institute for Biocomputation and Physics of Complex Systems (BIFI), University of Zaragoza, Zaragoza, Spain}
    \affiliation{Department of Theoretical Physics, Faculty of Sciences, University of Zaragoza, Zaragoza, Spain}
	
	\author{S. Boccaletti}
	\affiliation{Unmanned Systems Research Institute, Northwestern Polytechnical University, Xi'an 710072, China}
    \affiliation{Universidad Rey Juan Carlos, Calle Tulip\'an s/n, 28933   M\'ostoles, Madrid, Spain}
	\affiliation{Moscow Institute of Physics and Technology (National Research University), 9 Institutskiy per., Dolgoprudny, Moscow Region, 141701, Russian Federation}
	\affiliation{CNR - Institute of Complex Systems, Via Madonna del Piano 10, I-50019 Sesto Fiorentino, Italy}

\begin{abstract}
Elucidating the mechanisms that lead to the emergence, evolution, and survival of cooperation in natural systems is still one of the main scientific challenges of current times. During the last three decades, theoretical and computational models as well as experimental data have made it possible to unveil and explain, from an evolutionary perspective, key processes underlying the dynamics of cooperation. However, many common cooperative scenarios remain elusive and at odds with Darwin's natural selection theory. Here, we study evolutionary games on populations that are structured beyond pairwise interactions. Specifically, we introduce a completely new and general evolutionary approach that allows studying situations in which indirect interactions via a neighbor other than the direct pairwise connection (or via a group of neighbors), impacts the strategy of the focal player. To this end, we consider simplicial graphs that encode two- and three-body interactions.
Our simplicial game framework enables us to study the competition between all possible pairs of social dilemmas, and grants us the option to scrutinize the role of three-body interactions in all the observed phenomenology. Thus, we simultaneously investigate how social dilemmas with different Nash equilibria compete in simplicial structures and how such a competition is modulated by the unbalance of 2- and 1-simplices, which in its turn reflects the relative prevalence of pairwise or group interactions among the players.
We report a number of results that: (i) support that higher-order games allow for non-dominant strategists to emerge and coexist with dominant ones, a scenario that can't be explained by any pairwise schemes, no matter the network of contacts; (ii) characterize a novel transition from dominant defection to dominant cooperation as a function of the simplicial structure of the population; and (iii) demonstrate that 2-simplex interactions are a source of strategy diversity, i.e. increasing the relative prevalence of group interactions always promotes diverse strategic identities of individuals.  Our study constitutes, thus, a step forward in the quest for understanding the roots of cooperation and the mechanisms that sustain it in real world and social environments.
\end{abstract}

\maketitle


\section{Introduction}
\label{intro}
Cooperation is abundant and ubiquitous in natural systems, ranging from bacteria to human endeavours. Admittedly, our modern society is itself the result of thousands of years in which cooperative behavior has given rise to complex structures of relationships, norms, and in general to the possibility of coexistence despite the many differences between human beings. Moreover, our cooperative behavior has been shown to be key not only for the growth of our society, but also for the solution of many challenging troubles, such as  disease transmission \cite{bauch2004vaccination, wang2016statistical},  resource allocation \cite{shirado2019resource}, and other pressing challenges like climate changes \cite{vasconcelos2013bottom, wang2020communicating}.

The simplest form of cooperation involves two kind of strategists (or players): cooperators and defectors. A cooperator pays a given cost to allow individuals in the population to obtain a benefit, which is usually higher than the cost of cooperation \cite{West2007Social}. Defectors, on the other hand, are those individuals that exploit the situation by collecting the benefits produced by cooperators without paying costs. Although the emergence and sustainability of cooperation have been the subject of intense research in the last two decades, still many problems remain open, and a fundamental question is not yet fully answered: what are the mechanisms that give rise to cooperation? Significant advances in our understanding (and a partial answer to the previous question) were given in Ref. \cite{nowak_s06}, where Nowak individuated five mechanisms supporting cooperative behavior in nature: kin selection, direct reciprocity, indirect reciprocity, group selection and network reciprocity. In our work, we delve into the possible ways in which network reciprocity could enable cooperative behavior. It is worth stressing that network reciprocity, in which individuals are considered to interact following an underlying structure (a network), has been extensively studied theoretically  \cite{nowak1992evolutionary, qin2017social, matsuzawa2016spatial, santos_n08, santos_pnas06, perc_jrsi13}, but whether it plays a role or not in promoting cooperation remains still open to experimental validations \cite{cgl2012}.

Elucidating the mechanisms that promote cooperation is an important conceptual problem as well. The ubiquitous presence of cooperative behavior is compatible with Darwin's natural selection in some cases, but not always. For instance, cooperation due to kinship between individuals is a possible mechanism that as long as a cost is paid, contributes to propagate an individual's genes. Nonetheless, cooperation among unrelated individuals does not confer any additional fitness or selection advantage, and therefore individuals that bear the costs of cooperation should not become fixed in the population, and on the contrary they should go extinct after some generations. This would naturally lead to a population of all defectors, which, remarkably, is not what we observe in nature and our modern societies. Such an apparent contradiction is the main focus of evolutionary game theory (EGT) \cite{hofbauer1998evolutionary, nowak_06, sigmund_10, mesterton2019introduction}, whose ultimate goal is to explain how cooperative behavior emerges and unfolds in a plethora of systems.

\begin{figure}
	\centering
	\includegraphics[width=0.7\columnwidth]{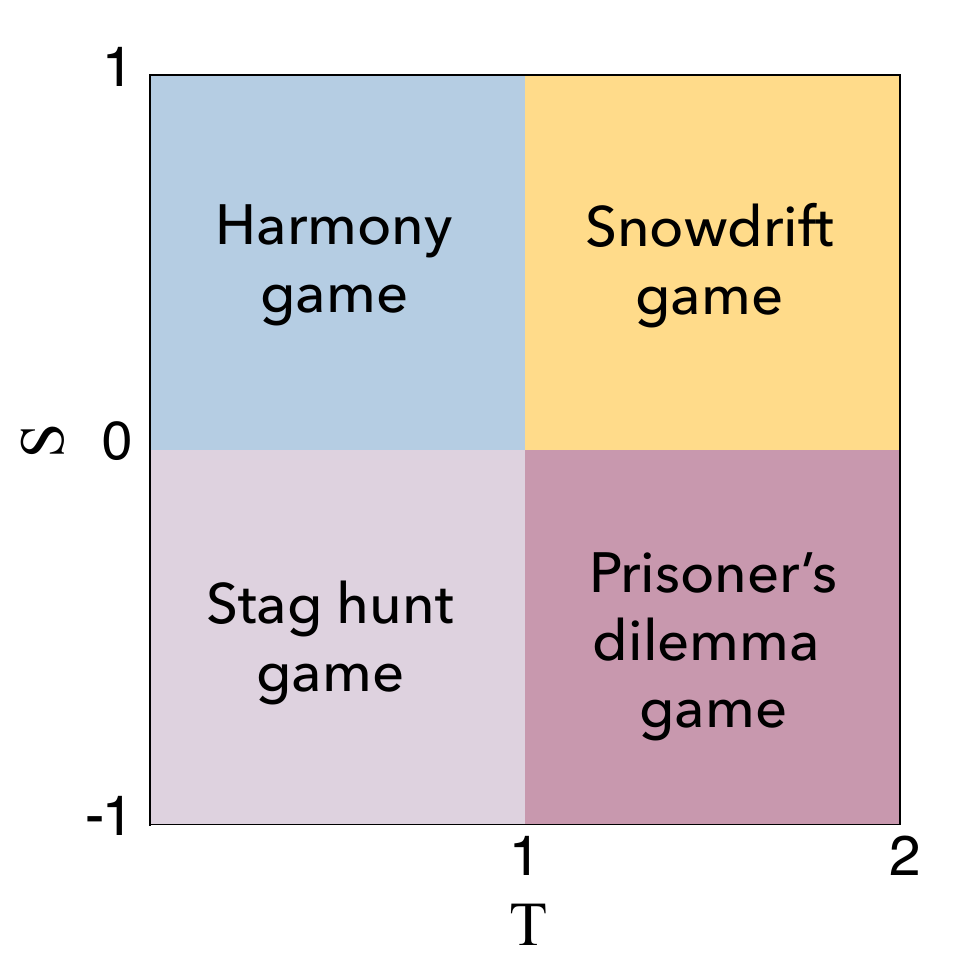}
	\caption{Schematic representation of all possible social dilemmas as a function of the pair of parameters $(T,S)$. The dilemmas shown in the quadrants have different Nash equilibria and fixed strategies, going from cooperation (Harmony Game) to defection (Prisoner's dilemma). $R$ and $P$ have been fixed to 1 and 0, respectively. See the text for further details.}
	\label{panelgame}
\end{figure}

In this paper, we follow the general methodology employed in EGT for the evolution of populations of strategies, but we implement it on more complex structures of interplays that are better to describe how individuals interact. Namely, the general framework of EGT is a two-agent two-strategy game \cite{szabo_pr07, perc_pr17, tanimoto_pre07, wang2015universal}, whereby each individual chooses a strategy from the set of the available ones [cooperate (C) or defect (D)] without knowing the strategy of its opponent. When two agents interact, a cooperator obtains a reward $R$ if interacting with a cooperator, and the so-called sucker’s payoff $S$ if interacting with a defector, whereas a defector gets $T$ (the so-called temptation to defect) if interacting with a cooperator, and a punishment $P$ if interacting with another defector. Different values of these parameters bring about a diversity of dilemmas which correspond to different equilibrium points, as we shall discuss in the next section.

Evolutionary models have allowed identifying additional mechanisms that could play a role to sustain cooperation among humans, including memory effects \cite{hilbe2017memory, Horv2012Limited}, strategy diversity \cite{sendina2020diverse, su2016interactive}, diverse forms of reputation \cite{Fu2008Reputation, gross2019rise} and aspiration \cite{wang2019evolutionary, hauser_n14} and onymity \cite{Wang-e1601444}. Central to our work, there have been many advances in the study of how interactions among humans are structured. These advances include the discovery and characterization of multilayer and interdependent networks \cite{kivela2014,boccaletti2014rev,aleta2019}, which have been shown to potentially lead to new forms of cooperative behavior \cite{jgg2012,wang_z_epl12, xia_njp18}. More recently, simplicial complexes and other forms of higher-order interactions \cite{Battiston2020,torres2020why} have also become amenable to a deeper scrutiny. The latter structures are of particular relevance because they allow to study situations in which individuals' interactions go beyond traditional pairwise connections in spatial and low-order networked evolutionary games. They include, for instance, group interactions, which are found more often than not (for instance, peer pressure effects on a given individual from a group of neighbors with which it is networking). Importantly enough, formulating evolutionary game dynamics in terms of higher-order models will allow to study scenarios in which pairwise and higher-order interactions coexist.

\begin{figure}
	\centering
	\includegraphics[width=7cm]{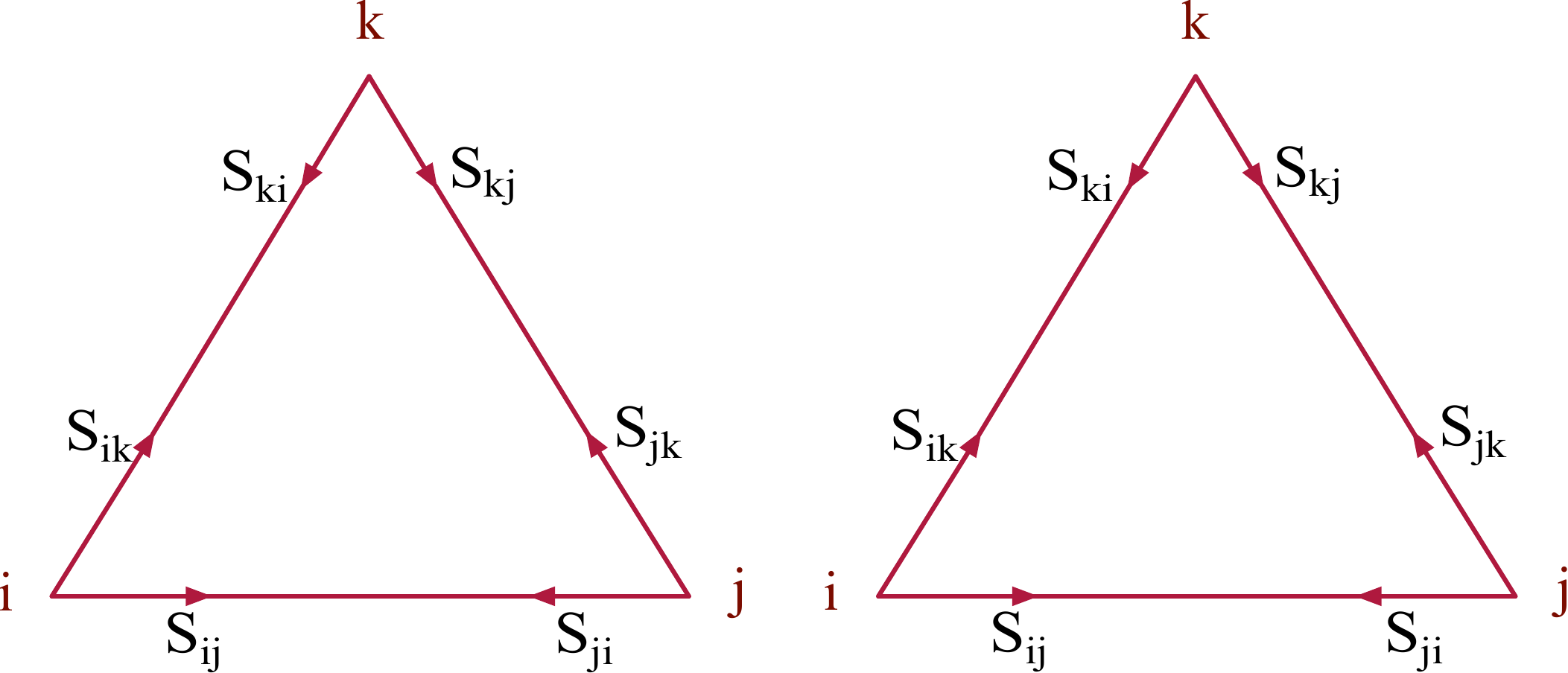}
	\caption{Sketches of a clique (left panel) and of a 2-simplex (right panel) formed by the 3 nodes $i$, $j$, and $k$. The quantity $s_{nm}$ represent the strategy that node $n$ adopts against node $m$.}
	\label{1dsimplex}
\end{figure}

When higher-order interactions are taken into account, the interplay between a given individual and one of its neighbors is not only affected by their two strategies, but also by that of one of its other neighbors, or even by those of a group of other neighbors. Therefore, it is a crucial problem that of investigating, as we are doing in this paper, simplicial game frameworks that explore the evolution of cooperative behavior as a function of the fraction of two-body and three-body interactions. In particular, simplicial complexes \cite{Battiston2020,torres2020why, gambuzza2020master} can effectively encode interactions between any number of units, including 0-simplex (a node), 1-simplex (a link), 2-simplex (a triangle), and so on. Our proposal considers the evolution of cooperative behavior combining 1-simplex (two-body) and 2-simplex (three-body) interactions. The latter three-body connections introduce new theoretical and algorithmic difficulties, since a common neighbor of two linked players could adopt either the same or different strategies with each of them. We study these scenarios and report  results of the phase diagrams of the evolutionary dynamics for different choices of the parameters that define the dilemmas, as well as results corresponding to the microscopic dynamics of the coexistence of cooperators and defectors. Next, we describe the evolutionary game model implemented here, including the definition of the simplicial game that allows us to study higher-order interactions.

\section{The Model for Simplicial Games}
\label{model}
\subsection{Generation of the simplicial structure}

We consider a network G composed of $N$ nodes, where the set of edges between nodes is uniquely coded into a symmetric $N \times N$ binary-valued adjacency matrix $A \equiv \{ a_{ij} \}$. This means that if an edge exists between nodes $i$ and $j$, then $a_{ij}$=1; otherwise, $a_{ij}$=0. Moreover, $k_i=\sum_{j=1}^N a_{ij}$ is the number of neighbors of player $i$, also called node $i$'s degree.
In order to implement and study the simplicial game, we first generate a substrate network following the rules proposed in Ref.\cite{kovalenko2020growing}, which include two possibilities: random and preferential connection rules.

We focus, in particular, on the case where the network is generated under the random scheme, which is as follows: (1) At $t=0$, generate a fully connected subgraph $G_1$ with a number of initial nodes $N_0$ (we use $N_0=5$); (2) At step $t=1$, add to the subgraph $G_1$ $m$ new nodes, which are linked to the two endpoints of $m$ edges that are randomly chosen among those already present in the subgraph $G_1$ (avoiding that the $m$ edges have overlapping nodes). Thus, the subgraph $G_1$ will have $m$ new triangles. In our study, we set $m=1$, which in its turn implies generating a final network with an average degree $\langle k \rangle=4$. Finally, (3) repeating step (2) until a network $G$ with $N$ nodes is formed. Using the previous algorithm, it is not difficult to check that the probability of each edge being selected in (2) is
\begin{equation}
	p=
	\frac {1}{((N_0 (N_0-1))/2+2m(t-1))}
\end{equation}

The preferential rule for connecting nodes consists of the same procedure as before except that when selecting an existing edge from the subgraph $G_1$ in step (2), this is done proportionally to the generalized degree of such an edge, that is, an edge $(ij)$ is selected with probability
\begin{equation}
\label{bellino}
	p=
	\frac {k_{ij}(t-1)}{\sum_{i,j}{k_{ij}(t-1)}} ,
\end{equation}
where $k_{ij}(t)$ is the generalized degree of edge $ij$ at time $(t)$, that is, the number of triangles formed (at time $t$) by the link $ij$.

It has to be highlighted that the non preferential (the preferential) case is encompassed by to the so-called {\it Network Geometry with Flavor} model \cite{Bianconi2015,Bianconi2016,Courtney2017} for the case of triangles, and flavor $s=0$ ($s=1$) in that model.
It is easy to demonstrate that the generated structure of links forms always (and only) triangles among the network's nodes.
Furthermore, Ref.\cite{kovalenko2020growing} demonstrated that the random scheme (which is the one used in the present Manuscript) generates a network with an imprinted highly heterogenous (power-law) distribution for the node degree, but with an associated rather homogeneous (exponentially decaying) distribution of the generalized degree.
In other words, while one can encounter great differences in the degree from a random choice to another of a network's node, different links of the graph participate essentially to the same number of triangles, with only small differences from a link to another.

In our study, we label each one of such triangles, and we introduce a parameter $0 \leq \rho \leq 1$
regulating the fraction ($1-\rho$) of triangular structures which are taken to be just cliques resulting from the closure of three 1-simplices, and the fraction ($\rho$) of triangles which are instead considered as genuine 2-simplex interactions. Namely, after generation of the network and labeling of all triangles, a fraction $\rho$ of randomly chosen triangles is taken to represent pure three-body interactions in the system, while the remaining fraction $1-\rho$ of triangles is considered to be the superposition of three links (i.e. three two-body interactions).

As for the evolutionary dynamics, we consider the family of symmetric 2x2 games, which is represented by the payoff matrix:
\begin{displaymath}
\bordermatrix{
	& C & D  	\cr
	C 	& 1 & $S$ 	\cr
	D	& $T$ & 0  	
}
\end{displaymath}
with only two free parameters, $S$ and $T$, that determine the equilibrium structures in the square ($-1 \le S \le 1$, $0 \le T \le 2$) of the $T-S$ space of the following four games: the Harmony game, the Prisoner's Dilemma game, the Snowdrift game and the Stag Hunt game (see Figure 1 for the parameter setting of these four dilemmas). Namely, the prisoner's dilemma verifies $S<0$ and $T>1$, and it has a unique strict Nash equilibrium corresponding to all defectors. The Stag Hunt game (or assurance game) is such that $S<0$ and $T<1$ and it has two pure Nash equilibria: CC and DD. The Snowdrift dilemma (or chicken game) corresponds to $T>1$ and $S<0$ and it has several Nash equilibria involving both C and D (this is an anti-coordination game). Finally, the Harmony game is defined for $T<1$ and $S>0$ and it represents a situation where mutual cooperation (CC) yields the maximum possible payoff to both players.

\begin{figure*}
	\centering
	\includegraphics[width=0.65\textwidth]{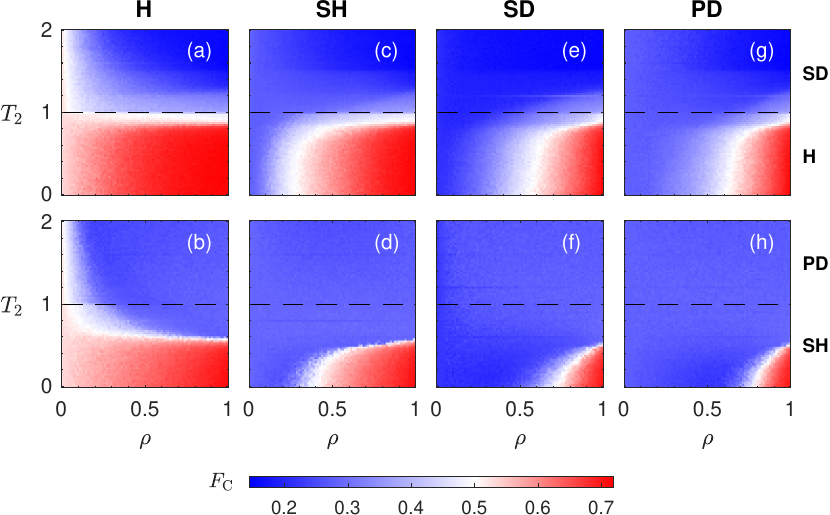}
	\caption{Contour plots of the cooperation frequency $F_C$ (the fraction of adopted cooperative strategies, see text for definition) in the asymptotic state, as a function of the fraction of 2-simplices $\rho$ and of $T_2$, for fixed values of $S_2$ (see color code reported in the bottom of the Figure). The other parameters are as follows: the first column corresponds to  $T_1=T_3 = 0.8$, $S_1=S_3 = 0.2$ which define the Harmony (H) game; the second column corresponds to $T_1=T_3=0.8$, $S_1=S_3=-0.2$ which define the Stag Hunt (SH) game; the third column corresponds to $T_1=T_3 = 1.2$, $S_1=S_3 =0.2$ which define the Snowdrift (SD) game; the fourth column corresponds to $T_1=T_3 = 1.2, S_1=S_3= -0.2$ which define the Prisoner’s Dilemma (PD) game. On the top of each column a proper label is displayed indicating Game1 and Game3. On the other hand, the values of $S_2$ also determine which dilemma corresponds to Game2. Specifically, we set $S_2=0.5$ for the first row of panels,  and  $S_2=-0.5$ for the second row of panels. All panels are furthermore divided horizontally by dashed lines positioned at $T_2=1$. Indeed, for the chosen setting, in panels a,c,e,g one has a SD game if $T_2\geq 1$ and a H dilemma for $T_2\leq 1$, whereas in panels b,d,f,h one has a SH dilemma if $T_2\leq 1$ and a PD game for $T_2\geq 1$. Proper labels are displayed on the right side of the figure indicating Game2. The substrate network has been generated starting from a fully connected graph with five nodes and adding one triangle (by means of the random procedure described in the text) at each time step, up to reaching a system size of $N=2,000$ nodes. $K=0.01$. The dynamics is evolved over 30,000 steps, and data refer to averages over the last 5,000 when the system is already settled into its asymptotic state. Furthermore, displayed data correspond to a single network realization, and a single initial condition.}
	\label{phase1}
\end{figure*}

\subsection{Definition of the simplicial game}
We start by defining a strategy matrix, $S=\{s_{ij}\}$, such that $s_{ij} = 1$ if player $i$ cooperates with player $j$, $s_{ij}=2$ if player $i$ defects when playing with $j$, and $s_{ij}=0$ if there is no connection between $i$ and $j$, i.e., when $a_{ij}=0$. $s_{ij}$ thus represents the strategy that node $i$ chooses when playing against node $j$. Consequently, a player $i$ is associated to a vector of independent strategies, the dimension of which is its degree $k_i$. In other words, in a particular instance of a game, players can simultaneously cooperate with some of their neighbors and defect with others. Thus, $k_i=k_i^C + k_i^D$, where $k_i^C=\sum_{j|s_{ij}=1}a_{ij}$ ($k_i^D=\sum_{j|s_{ij}=2}a_{ij}$) is the time-dependent number of neighbors player $i$ is currently cooperating (defecting) with.

Now, definition of an evolutionary game on a structure of pairwise interactions (a network) implies the introduction of a payoff matrix. In our case, instead, defining a 2-simplicial game implies the definition of a payoff tensor, which is the object of the next sub-section.

\subsection{Calculation of the payoff on 1-simplices and on 2-simplices}
Let us postulate that the payoff $\Pi_i$ of node $i$ is accumulated on each link, and let us illustrate how payoffs are earned by considering the payoff of node $i$ calculated for the specific link $(i,j)$. The link participates to $k_{ij}$ triangles, and therefore the accumulated payoff of node $i$ in that link is:
%
\begin{equation}
\Pi_{i,(ij)} =\frac{1}{k_{ij}} \sum_{\tau \in \bigtriangleup} \Pi_{i,(ij),\tau},
\end{equation}
where the sum runs over all the elements $\tau$ of the set $\bigtriangleup$ which contains all the $k_{ij}$ triangles formed by the link $ij$, $\Pi_{i,(ij)}$ is the accumulated payoff of node $i$ along the specific link $(ij)$, and  $\Pi_{i,(ij),\tau}$  is the payoff of node $i$ along the specific link $(ij)$ with respect to the specific triangle $\tau$.

Now, it is crucial to distinguish between the case in which a given triangle $\tau$ represents just the sum of three 1-simplices (i.e., it is a triangle formed by the closure of three separate links, as depicted in the left sketch of Fig.~\ref{1dsimplex}) and the case in which the triangle $\tau$ stands instead for a three-body interaction, namely, a 2-simplex as illustrated in the right sketch of  Fig.~\ref{1dsimplex}.

\begin{figure*}
	\centering
	\includegraphics[width=0.65\textwidth]{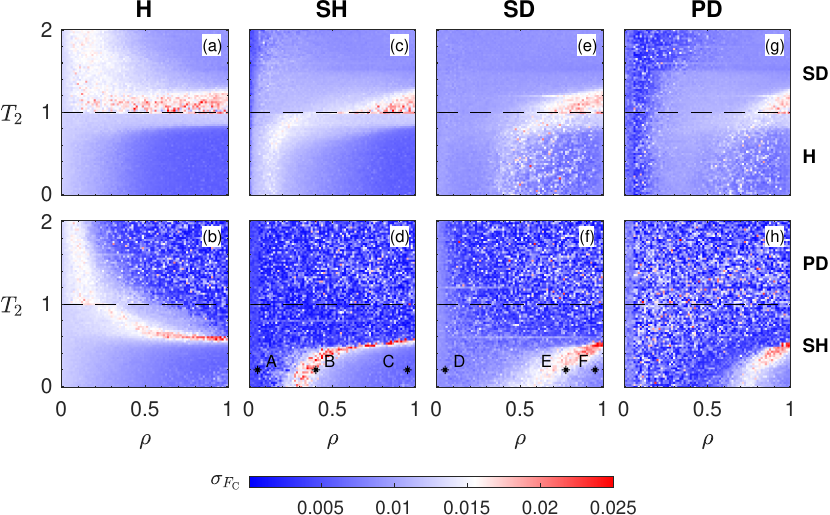}
	\caption{Standard deviation $\sigma_{F_C}$ of the cooperation frequency depicted in Fig.~\ref{phase1} (see text for definition) as a function of $\rho$ and $T_2$. The color code is reported at the bottom of the Figure.  Notice that in the regimes where one of the two strategies, cooperation or defection, dominates, the standard deviation almost vanishes, whereas it is larger and larger when more and more coexistence of the two strategies occurs. All parameter values and stipulations as in the caption of Fig.~\ref{phase1}.  In panels d and f, stars indicate the locations in the parameter space where the time dependent plots of Fig.\ref{timeevolution} are reported, with labeling letters indicating the specific panel of Fig.\ref{timeevolution} they are referring to. All labeled star corresponds to $T_2=0.2$. The other parameter is: $\rho=0.05$ (A and D), $\rho=0.95$ (C and F), $\rho=0.4$ (B), and $\rho=0.775$ (E).}
	\label{STD1}
\end{figure*}

If the triangle $\tau$ is the closure of three 1-simplices,  then in this case the link $(ij)$ just represents a pairwise interaction between nodes $i$ and $j$, and the value of $\Pi_{i,(ij),\tau}$ in that triangle will be simply obtained from a payoff matrix
\begin{displaymath}
\bordermatrix{
	& C & D  	\cr
	C 	& 1 & S_1 	\cr
	D	& T_1 & 0  	
}.
\end{displaymath}
We henceforth refer to this sort of interactions as Game1.

When instead the triangle $\tau$ stands for a 3 body-interaction, it means that the link $(ij)$ is part of a 2-simplex (see the right sketch of Fig.~\ref{1dsimplex}).
In this case, therefore, the computation of $\Pi_{i,(ij),\tau}$ needs to involve explicitly also the strategic state of node $k$ which is closing the 3 body-interaction with nodes $i$ and $j$. This implies that we need to introduce a tensor. The procedure is as follows:

1. Check the strategies $s_{ki}$ and $s_{kj}$ that node $k$ is using against nodes $i$ and $j$.

2. If $s_{ki}=s_{kj}$, then, nodes $i$ and $j$ play Game2 along link $(ij)$, i.e., the payoff is calculated from the matrix
\begin{displaymath}
\bordermatrix{
	& C & D  	\cr
	C 	& 1 & S_2 	\cr
	D	& T_2 & 0  	
}
\end{displaymath}

3. If $s_{ki} \neq s_{kj}$, then, nodes $i$ and $j$ play Game3 along the link $(ij)$, i.e., the payoff is calculated using the matrix
\begin{displaymath}
\bordermatrix{
	& C & D  	\cr
	C 	& 1 & S_3 	\cr
	D	& T_3 & 0  	
}
\end{displaymath}

\begin{figure*}
	\centering
	\includegraphics[width=0.7\textwidth]{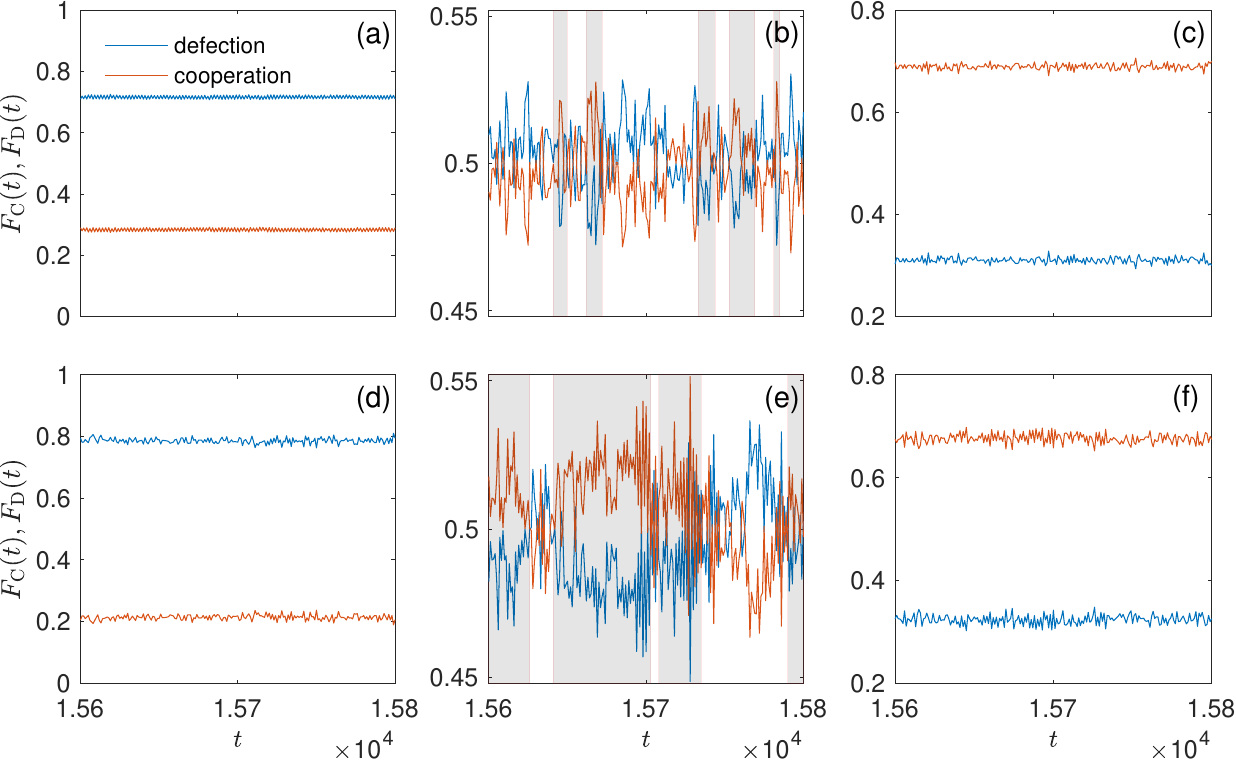}
	\caption{Time evolution of cooperators' ($F_C(t)$) and defectors' ($F_D(t)$) frequencies for different values of the pair $(T_2,S_2)$ and three values of $\rho$. See text for definitions. The first row of panels corresponds to the case in which Games 1 and 3 are fixed to a SH game, and compete with another SH game as Game 2. The letter labeling each panel corresponds to the parameter choice indicated with a star and labeled with the same capital letter in Fig. \ref{STD1}. The second row of panels corresponds instead to the case in which Games 1 and 3 are fixed to a SD game, and compete with a SH game as Game 2. Also in this case, the letter labeling each panel corresponds to the parameter choice indicated with a star and labeled with the same capital letter in Fig. \ref{STD1}. In both  upper and lower rows of panels one can clearly see the emergence of two distinct regimes: a first one where one of the two strategies is dominant over the other (panels a, c, d, f), and a second one where instead there is a time-dependent coexistence of cooperators and defectors (panels b and e), with dominant strategies alternating in time. Shaded gray areas in panels b and e mark the time intervals in which cooperators dominate over defectors. See the captions of Figs. \ref{phase1} and \ref{STD1} for all other parameters and stipulations.}
	\label{timeevolution}
\end{figure*}

It is worth stressing that the above rules imply that a node could be playing, concurrently, different dilemmas with different neighbors or group of neighbors depending on the parameterizations of the payoff matrices of Game1, Game2 and Game3. Notice that the conditions 1-3 above can be chosen also in other ways to actually define a 2x2x2 payoff tensor (the superposition of the two matrices defining Games 2 and 3). The motivation of our choice is dictated by the fact that in social endeavours one individual may behave differently in its relationships with a neighbor if he sees that a third individual is or isn't treating the two of them in the same footing.

An evolutionary step finishes when the payoffs of all nodes are calculated. The evolution of the population of strategists proceeds using a Fermi rule. Specifically, the total payoff of node $i$ is calculated as
\begin{equation}
\ \Pi_i = \frac{1}{k_i}\sum_{j \in N_{i}} \Pi_{i,(ij)}
\end{equation}
where $N_i$ represents the neighbors of node $i$. Next, each node imitates the strategy adopted against it by its neighbor $\tilde j$ which accumulated the highest total payoff in the current step, i.e., node $i$ updates its $k_i$ strategies with probability
\begin{equation}
W=
\frac{1}{1+e^{\left[\left(\Pi_i-\Pi_{\tilde j}\right) / K\right]}}. \label{fermi}
\end{equation}

In our simulations, we set (unless otherwise specified) the so-called Fermi temperature $K$ equal to 0.01, and the strategies are synchronously updated, that is, all the agents update their strategy vector at the same time. In what follows, we start by showing the emerging scenario when one varies the proportion, $\rho$, of three-body interactions in the underlying network, and inspects this way the impact of having higher-order interactions on the evolutionary dynamics.

\section{Results and Discussions}
\label{results}

\subsection{Competition of social dilemmas}

Calibrating the parameters of Game1 and Game3 to be the same (i.e., setting $T_1=T_3$ and $S_1=S_3$) and varying those of Game2 ($T_2$ and $S_2$), our simplicial game framework enables us to study the competition between all possible pairs of social dilemmas.
Furthermore, increasing $\rho$ grants to probe and scrutinize the role of three-body interactions in all the observed phenomenology. This way one can simultaneously investigate how social dilemma with different Nash equilibria compete in simplicial structures and how such a competition is modulated by the unbalance of 2- and 1-simplices, which in its turn reflects the relative prevalence of pairwise or group interactions.

The results are shown in Fig.~\ref{phase1} where, from the first to the fourth columns, Game1 and Game3 are assigned to, respectively, the Harmony (H), Stag Hunt (SH), Snowdrift (SD) and Prisoner's Dilemma (PD) games.  Figure~\ref{phase1} reports the $T_2 - \rho$ phase diagram of the (average) frequency of cooperation set in the asymptotic game dynamics, i.e. the time average of the overall fraction of cooperative strategies. Furthermore, we have set $S_2=0.5$ in the first row of panels,  which corresponds to Game2 being an Harmony dilemma if $T_2 \leq 1$ or a SD game when $T_2 \geq 1$. Likewise, when $S_2 = -0.5$ (second row of panels in Fig.~\ref{phase1}), Game2 becomes a SH game if $T_2 \leq 1$, and a PD for $T_2 \geq 1$. In this way, one obtains the emergent dynamics for all possible competitions between pairs of dilemmas in the system.

Let us now discuss the observed scenarios in Fig.~\ref{phase1}, paying particular attention to the impact of higher order games. For simplicity, we start by analyzing the results shown in the panels of the first column, which correspond to Game1 and Game3 both being Harmony games. Since the equilibrium point for this case is $CC$, we should observe that the dominant strategy for low values of $\rho$ $-$ that is, regardless of what is represented by Game2$-$, is cooperation. This is indeed what one obtains. As $\rho$ increases, however, so does the number of 2-simplices in the system and thus the likelihood that a given node $i$ is involved in different games. In such a case, the cumulative payoff starts to depend more strongly on the results of the dynamics of Game2, which can be either of the 4 dilemmas. When $T_2\leq 1$ no matter whether $S_2>0$ or $S_2<0$, the equilibria of Game2 tend to favor cooperation (they are H and SH games, respectively), and thus, the effect of increasing $\rho$ is not noticeable. On the contrary, when $T_2\geq 1$, Game2 represents either a SD dilemma ($S_2>0$) or a PD game ($S_2<0$), both of which are detrimental to coordination or cooperation. Therefore, as $\rho$ increases, the fraction of cooperative behavior decreases.

A richer, and remarkable, scenario emerges when Game 1 and Game 3 are set to be SH, SD and PD (i.e. in the second, third and fourth columns of Fig.~\ref{phase1}). Now, for pairwise interactions ($\rho=0$) one has the setting of a dominant defective state. However, when $T_2<1$ (i.e., when Game 2 is either the Harmony Game, as in the lower half-panels of the first row, or the Stag Hunt Game, as in the lower half-panels of the second row) a clear transition occurs, as $\rho$ increases, toward dominant cooperation.
Such a transition is, therefore, fully due to the prevalence of 2-simplex interactions in our system. The transition occurs at all $T_2<1$ for the Harmony Game, which is not so surprising, given the fact that the Nash equilibria of the H game is full cooperation: increasing $\rho$ (and therefore increasing the number of times the system plays the H game) one should expect a transition to prevalent cooperation.
The absolutely non trivial case is when Game 2 is set to be SH, as this game has two pure Nash equilibria: CC and DD. In this case, Fig.~\ref{phase1} shows that for $T_2< \tilde T_2 <1$ a transition still occurs toward complete cooperation, even for the case for which Games 1 and 3 are in PD (see panel h).

The conclusion is that for $T_2<1$ a regime is always found such that increasing the prevalence of three-body interactions in our network, the competition of the two dilemmas conduces to the existence and maintenance of cooperation.
In other words, the impact of higher-order games for the evolution of cooperation radically changes when the dominant strategy is defection. In these cases,  when $\rho$ increases, cooperators have a chance to invade an otherwise fully defectors population and survive for large enough values of $\rho$ when $T_2<1$, i.e., when Game2 is either a H dilemma ($S_2>0$) or a SH game ($S_2<0$).

\subsection{Transition between the two strategies regulated by 2-simplex interactions}

The observed transitions are associated to consistent fluctuations in time of the strategies of each player.
Further to the previous analysis, and in order to examine such microscopic traits in the dynamics, we have therefore monitored the volatility of both cooperation and defection, measuring the standard deviation of the frequencies of each strategy. Specifically, we compute
\begin{equation}
	F_C(t) = \frac{\sum_{i}F_{C_i}(t)}{N}
\end{equation}
 and
\begin{equation}
F_C = \langle F_C(t)\rangle_T  ,
\end{equation}
where $F_C(t)$ represents the total cooperation rate in the network at step $t$, and $F_C$ and $\sigma_{F_C}$ are, respectively, the time-averaged cooperation density over an observation time $T$ and the standard deviation of $F_C(t)$. Trivially, $\sigma_{F_C}$ equals to 0 if the evolution falls into a frozen (absorbing) state, and larger than 0 if the system settles into an asymptotic state in which the fraction of strategists is time dependent and fluctuates around an average value $-$ the more intense the fluctuations, the larger the values of $\sigma_{F_C}$. The results are shown in Fig.~\ref{STD1}: $\sigma$ drops to 0 when either cooperators or defectors dominate the network, but increases above zero when cooperators and defectors coexist, reaching its maximum value exactly at the transition areas from one to another dominant strategy.
The latter happens whenever the equilibrium corresponding to a dominant strategy destabilizes due to the increase of the number of triangles and therefore there is a bigger impact of Game2 interactions on the outcomes of the dynamics, as it can be clearly seen when comparing Fig.~\ref{phase1} with Fig.~\ref{STD1}.

To further explore the evolutionary dynamics of the populations of the two strategists, we chose six scenarios for a closer inspection taken from the competition of SH and SD (as Games 1 and 3) with SH as Game 2.
In Fig.~\ref{timeevolution}, we report the evolution of cooperation and defection strategies in the six scenarios.
The Figure clearly shows that the overall simplicial game evolves to a state in which either one of the two strategies dominates over the others with only residual fluctuations (as in panels a, c, d and f), or to a time-dependent asymptotic state where the system enters into a cycle of alternate dominance of cooperation and defection (as in panels b and e), with an average overall frequency of either of the two strategies equal to 0.5. In this latter regime, fluctuations also intensify and actually determine, time by time, which of the two strategies is majority in the system (the shaded gray areas in panels b and e mark actually time intervals in which cooperators dominate over defectors).

\begin{figure}
  \centering
  \includegraphics[width=\columnwidth]{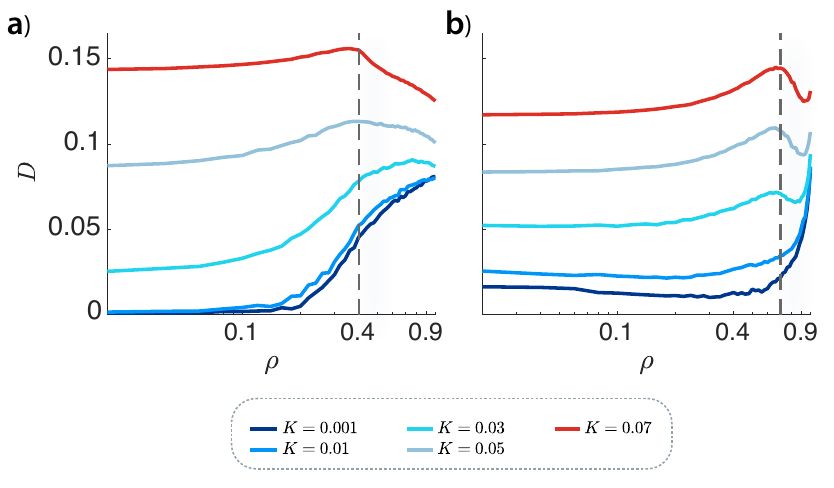}
  \caption{Network strategy diversity $D$ (see text for definition) vs. the fraction $\rho$ of 3-body interactions for increasing values of the temperature along (a)  the line passing through points A-C in Fig.~\ref{STD1}(d), and (b) that containing the points D-F in Fig.~\ref{STD1}(f). The horizontal axis is in logarithmic scale, and the color code for the different curves in both panels is reported in the legend at the bottom of the figure. In all cases, increasing the fraction of 2-simplices induces a substantial increase in the diversity of the players. Vertical lines are drawn at the points where fluctuations of the cooperation frequency are maximal in Fig.~\ref{STD1}(d) and (f). \label{fig6}}
\end{figure}

\subsection{2-simplex interactions induce players' diversity}

In order to further investigate on the microscopic features of the observed fluctuations, we calculate here the link strategy index,
defined by
$$\ell_i=1-\frac{|{k_i^D-k_i^C}|}{k_i}.$$
Such an index quantifies the {\it rigidity} of each node's strategy vector: it indeed vanishes for all those players which display strategy vectors made of all cooperation or all defection entries (i.e. those players which actually adopt a unique node strategy against all their neighbors), whereas it gets larger and larger the more diversified the players' identities are.

Once the link strategy index has been calculated for each node of the network,  a network strategy diversity $D$ can be defined in the asymptotic state (i.e. at a final time $t_f$) by just averaging over time and over all nodes:
$$ D = \frac{1}{NT}\sum_{i=1}^{N} \sum_{t_k=t_{f}-T}^{t_{f}}   \ell_i (t_k), $$
where the node average runs over all the elements of the network, and the time average runs over the last $T$ steps of the system's evolution.

Figure \ref{fig6} reports the values of $D$ that correspond to increasing fractions $\rho$ of 3-body interactions and increasing values of the temperature $K$, calculated along  the line passing through points A-C in Fig.~\ref{STD1}(d) (left panel) and that containing the points D-F in Fig.~\ref{STD1}(f) (right panel). It is seen that, at any fixed value of $\rho$, $D$ increases with $K$, which is not surprising as rising the temperature has the consequence of rocketing noise effects in Eq. (\ref{fermi}), which in its turn regulates the way all strategies are updated.
What is instead remarkable is that at all fixed temperatures (even at very low ones, as in the dark blue line of the left panel) $D$ substantially increases with $\rho$, indicating that 2-simplex interactions are actually a source of diversity in the network.
By comparing Fig. \ref{fig6} and Fig.~\ref{STD1},  one furthermore sees that at higher temperatures $D(\rho)$ displays a maximum where fluctuations of the cooperation frequency are maximal.

\section{Conclusions}
\label{discuss}

Altruism is the act of benefiting others at the expenses of one's own interests. Cooperation maintains social stability, but it is not always compatible with
Darwin's natural selection mechanisms. For instance, cooperation due to kinship between individuals may contribute to propagate an individual's genes, but cooperation among unrelated individuals does not confer any selection advantage, and therefore individuals that bear the costs of cooperation should go extinct after some generations, which is not instead what one observes in nature and modern societies.

In our study, we revealed some novel mechanisms which could be at the basis of the emergence and maintenance of cooperation in a networked population.
Namely, we considered evolutionary game theory, and showed how to implement it on more complex structures of interplays, such as simplicial complexes, where pairwise and higher-order interactions coexist.
The framework we introduced enables one to simultaneously investigate how social dilemmas with different Nash equilibria compete in simplicial structures and how such a competition is modulated by the unbalance of 2- and 1-simplices, which in its turn reflects the relative prevalence of pairwise or group interactions.

A series of novel, and remarkable, results are found.

First of all, it is seen that increasing the prevalence of three-body interactions, the competition of dilemmas conduces to the existence and maintenance of cooperation.
In other words, higher-order games allow for non-dominant strategists to emerge and coexist with dominant ones, eventually taking over the dynamics in some parameter regions.
Therefore, our results provide an explanation, based on higher-order interactions, for situations in which cooperation prevails despite the fact that evolutionary game dynamics in well-mixed or networked populations would not support it.

A second result is that the transition from dominant defection to dominant cooperation (as the number of 2-simplex interactions increases) is characterized by fluctuations in time which display a maximum exactly at the transition point. In practice, the system sets either on a state where one of the two strategies asymptotically dominates over the others at an almost constant value with only residual fluctuations, or on a time-dependent asymptotic state where dominance of cooperation and defection alternates, with an average overall frequency of either of the two strategies equal to 0.5, and with intensified fluctuations that actually determine, time by time, which of the two strategies is majority in the system. This latter scenario resembles the outcome of processes implying dichotomic choices in modern societies, such as, for instance, the presidential US elections. There, indeed, one has essentially a bipolar system where two parties compete regularly for the presidency and which sees two main blocks of voters forming the electoral core of the two parties (two clusters of singleton strategists) and a consistent group of swing voters whose choice is the one which actually determines, time by time, the final outcome, i.e. the prevalence of a party over the other.

The third novel result is that 2-simplex interactions are a source of strategy diversity in the network, i.e. increasing the relative prevalence of group interactions always promotes diverse strategic identities of individuals.
Strategy diversity is of particular importance in the evolutionary dynamics of structured populations \cite{sendina2020diverse, su2016interactive}, as it overcomes the limit of node's strategies, which certainly can be a reasonable choice for organisms with no or limited self-awareness and intelligence, but which becomes unrealistic for more complex living beings, as humans and many other animals act differently with certain peers than they do around others.
Moreover, it is known that monotonic strategies lead, in general, to time stationary network's arrangements, i.e. the setting (from a given time on) of a population of {\it simpletons} where the identity of each unit is that of a permanent cooperator or of a permanent defector. This is also far from properly representing real interactions in human or animal societies, wherein members actually alternate cooperation and defection in time, in a way that is very much similar to that displayed by us in Figure ~\ref{timeevolution}.  In social systems, for instance, the human propensity to cooperation does vary in time, and this has important consequences for the outcomes in decision conflicts \cite{Evans2015}, and competitive environments \cite{Cone2014}. In biophysical systems, changes in cooperative behavior may be observed over time due to population feedbacks \cite{Sanchez2013}, or to varying resource availability \cite{Hoek2016}.

Finally, it is important to mention that the shown results are robust with respect to network's size and average degree, and to external temperatures, in the sense that qualitatively similar behaviors can be obtained by changing $N$, $\langle k \rangle$ and $K$ within rather extended ranges of their values. The case where also generalized degrees are heterogeneous [which would imply the use of Eq. (\ref{bellino}) in the graph's generation process] will be presented elsewhere.

\section*{Acknowledgements}
\label{ackn}

I.S.N. acknowledges partial support from the Ministerio de Econom\'ia, Industria y Competitividad of Spain under project FIS2017-84151-P. Y.M. acknowledges partial support from Intesa Sanpaolo Innovation Center, the Government of Arag\'on and FEDER funds, Spain through grant ER36-20R to FENOL, and by MINECO and FEDER funds (grant FIS2017-87519-P). S.B. acknowledges partial support from the Project EXPLICS of the Italian Ministry of Foreign Affairs.

The funders had no role in study design, data collection, and analysis, decision to publish, or preparation of the manuscript.


\end{document}